\documentclass[epj]{svjour}
\usepackage{graphicx}
\begin{document}
\title{Flow to strong coupling in the two-dimensional Hubbard model}
\author{Carsten Honerkamp$^{1}$\thanks{New address: Department of Physics, Massachusetts
    Institute of Technology, Cambridge MA 02139, USA,
    \email{carsten@mit.edu}}, Manfred Salmhofer$^{2}$, and T.M.
  Rice$^{1}$} \institute{$^{1}$ Theoretische Physik,
  ETH-H\"onggerberg, CH-8093 Z\"urich, Switzerland \\ $^{2}$
  Theoretische Physik, Universit\"at Leipzig, D-04109 Leipzig,
  Germany}

\date{March 13, 2002}

\abstract{ We extend the analysis of the renormalization group flow in
  the two-dimensional Hubbard model close to 
  half-filling using the recently developed temperature flow
  formalism.  We investigate the interplay of $d$-density wave and
  Fermi surface deformation tendencies with those towards $d$-wave
  pairing and antiferromagnetism. For a ratio of next nearest to
  nearest neighbor hoppings, $t'/t=-0.25$, and band fillings where the
  Fermi surface is inside the Umklapp surface, only the $d$-pairing
  susceptibility diverges at low temperatures.  When the Fermi surface
  intersects the Umklapp surface close to the saddle points, $d$-wave
  pairing, $d$-density wave, antiferromagnetic and, to a weaker
  extent, $d$-wave Fermi surface deformation susceptibilities grow
  together when the interactions flow to strong coupling. We interpret
  these findings as indications for a non-trivial strongly coupled
  phase with short-ranged superconducting and antiferromagnetic
  correlations, in close analogy with the spin liquid ground state in
  the well-understood two-leg Hubbard ladder.   } \authorrunning{Honerkamp,
  Salmhofer, and Rice} \titlerunning{Flow to strong coupling in the 2D
  Hubbard model} \maketitle
\section{Introduction}
The two-dimensional (2D) Hubbard model is a focus of electron
correlation theory because of its relevance to high-temperature
superconductivity~\cite{pwa} and other correlation phenomena such as
metal-insulator transitions~\cite{gebhard} and itinerant
ferromagnetism~\cite{vollhardt}. Apart from proposed exotic
spin-liquid or non-Fermi liquid states~\cite{pwa,pwacondmat,varma} a
widely adopted approach is to classify possible ground states of the
model in terms of broken symmetries. This can be done for the
strong-coupling versions of the Hubbard model by invoking spin-charge
separation~\cite{u1} or large-spin approaches~\cite{sachdev}. But for
moderate interaction strengths it is more common to invoke a mean
field description of long-range ordered ground states with anomalous
expectation values which are bilinear in electron creation and
annihilation operators.  These theories are either constructed with
the bare Hubbard interaction~\cite{linhirsch,hofstetter} or with an
effective interaction which is postulated or calculated
perturbatively, e.g. by one-loop renormalization group (RG) treatments
or parquet summations.  Typically these treatments are restricted to
the weakly correlated case, but often they can give qualitatively
satisfactory results such as the antiferromagnetic state at
half-filling or a $d_{x^2-y^2}$-superconducting ground state at a
reduced band filling in the 2D Hubbard
model~\cite{zanchi,halboth,honerkamp,honerkamped,irkhin}.

Recently two independent proposals enlarging the variety of potential
symmetry-broken states have been made. One possibility was raised by
Halboth and Metzner~\cite{halboth}, based on an analysis of the Landau
function extracted from a RG flow. They pointed out that the system
might become unstable against a Pomeranchuk instability, i.e. a
spontaneous deformation of the Fermi Surface (FS) reducing its
symmetry from square to orthorhombic. Qualitatively similar results
were obtained by Valenzuela and Vozmediano\cite{valenzuela} in a
mean-field study of an extended Hubbard model. However with this
technique and also with the Wilsonian RG formalism based on an
infrared (IR) cutoff used in Ref.~\cite{halboth} it is difficult to
compare the strength of the FS deformation tendencies with other
potential singularities, e.g. in the AF or $d$-wave superconducting
channel.  Here we use the recently developed temperature-flow
formalism~\cite{honerkampfm}, to analyze these FS deformation
tendencies together with other channels in an unbiased way.

A different type of symmetry breaking was considered by
Nayak~\cite{nayak} and Chakravarty et al.~\cite{chakravarty}. They
proposed that a $d$-density-wave state (closely related to flux
phases~\cite{marston} or orbital antiferromagnets~\cite{schulz}) forms
at a high temperature in the underdoped high-$T_c$ cuprates and is
responsible for the many anomalous properties like the notorious
pseudogap.  This proposal has been controversial for a number of
reasons\cite{dhlee,palee}, not least because of the absence of experimental
evidence that the onset of the pseudogap is accompanied by an actual
phase transition rather than a continuous crossover. The onset of a
phase with broken translational and time reversal symmetries should
cause anomalies in thermodynamic and other properties which have not
been observed.  Nonetheless it is interesting to investigate whether
$d$-density-wave tendencies play a significant role in the RG flow to
strong coupling.

In this paper we show that in the one-loop RG tem\-perature-flow to
strong coupling in the $t$-$t'$ Hubbard model at $U \approx 3t$ both
the FS deformation and $d$-density wave tendencies grow. However their
growth is weaker than that of $d$-wave pairing and AF fluctuations, so
that they do not appear as the dominant instabilities of the model.
Nonetheless, similar to our findings with a RG IR-flow in
Ref.~\cite{honerkamp}, we find a particular density region - the
so-called {\em saddle-point regime} which occurs when the FS is close
to the saddle points of the band dispersion at $(\pi, 0)$ and $(0,
\pi)$ - where $d$-wave pairing, AF, $d$-density wave and $d$-wave
Pomeranchuk fluctuations are intimately related and grow together in
the range where the one-loop flow is credible. This contrasts with the
behavior for fillings smaller than the van Hove filling where only the
$d$-wave pairing correlations continue to grow towards lower
temperatures, while the others are cut off below some temperature.

The qualitative picture obtained here is a confirmation and extension
of the results described in Ref.~\cite{honerkamp}.  The main
motivation for the present paper is twofold: {\em a)} we show that the
new temperature-flow RG scheme reproduces the results of the IR-flow
RG ~\cite{honerkamp} and yields an even clearer picture; {\em b)} we
now also include $d$-wave FS deformations and $d$-density wave
tendencies which were not considered in the previous analysis.

In the following we first the describe the calculational scheme and
the symmetry-breaking tendencies which are investigated. Then we
discuss our numerical results for the case of zero and non-zero values
of the next-nearest hopping $t'$ and conclude with a comparison of our
observations to ladder systems, which we argue give some insights into
the nature of the strongly coupled state.

\section{The calculational scheme}
The Hamiltonian for the 2D $t$-$t'$ Hubbard model is 
\[ H= -t \sum_{\mathrm{n.n.} , \, s } c^\dagger_{i,s} c_{j,s} -t'
\sum_{\mathrm{n.n.n.},\, s } c^\dagger_{i,s} c_{j,s} +U \sum_i
n_{i\uparrow} n_{i \downarrow} \, \] with onsite repulsion $U$ and
hopping amplitudes $t$ and $t'$ between nearest neighbors (n.n.) and
next-nearest neighbors (n.n.n.) on the 2D square lattice.  We apply
the so-called temperature-flow RG scheme introduced
recently~\cite{honerkampfm} in a $N$-patch implementation that covers
the full Fermi surface .  Similar to the approaches in
Refs.~\cite{zanchi,halboth,honerkamp,salmhofer} the $T$-flow scheme is
derived from an exact RG equation. However, a low energy IR cutoff is
not used, and instead the temperature $T$ itself is used as the flow
parameter. This new formulation allows an unbiased comparison between
AF and FM tendencies, whereas RG schemes with a flowing IR cutoff
artificially suppress particle-hole excitations with small
wave vectors, e.g. long wavelength density fluctuations.  For a
detailed discussion of this matter and a derivation of the $T$-flow RG
equations the reader is referred to Ref.~\cite{honerkampfm}. 

The RG formalism yields a hierarchy of differential equations for the
one-particle irreducible $n$-point vertex functions, $\Gamma_T^{(n)}$,
as functions of the temperature. Integration of this system of
equations gives the temperature-flow. As initial condition we assume
that at a high temperature $T_0$, the single-particle Green's function
of the system is simply $G_0 (i \omega , \vec{k} ) = [ i \omega -
\epsilon (\vec{k}) ]^{-1} $ and the interaction vertex is given by a
local repulsion, $U$.  This is justified if $T_0$ is sufficiently
large, as wavevector-dependent perturbative corrections to selfenergy 
and four-point vertex  decay at least $\propto 1/T_0$\footnote{A wavevector-independent contribution can be absorbed into the chemical potential $\mu$. Flows with fixed particle number instead of fixed $\mu$ give qualitatively similar results.}.  
We truncate the infinite system of equations by dropping all
vertex functions $\Gamma_T^{(n)}$ with $n>4$.  In the present
treatment we also neglect selfenergy corrections and the frequency
dependence of the vertex functions.  This restricts the scheme to
one--loop equations for the spin-rotation invariant four-point vertex
$\Gamma_T^{(4)}$.  Starting with weak to moderate interactions, we
follow the $T$-flow of $\Gamma_T^{(4)}$ as $T$ decreases.

The four-point vertex $\Gamma_T^{(4)}$ is determined by a coupling
function $V_{T} (\vec{k}_1, \vec{k}_2, $ $\vec{k}_3)\;$ (see
Refs.~\cite{honerkamp,honerkampfm,salmhofer}).  The numerical
implementation follows the work of Zanchi and Schulz~\cite{zanchi} and
was already explained in Refs.~\cite{honerkamp} and \cite{honerkampfm}.
We define elongated phase space patches around 
lines leading from the origin of the BZ to the $(\pm \pi,\pm
\pi)$-points, and approximate $V_{T} (\vec{k}_1,\vec{k}_2,\vec{k}_3)$
by a constant for all wave vectors in the same patch.  We calculate
the RG flow for the discrete subset of interaction vertices with each
FS patch represented by a single wave vector and with the initial
condition $V_{T_0} (\vec{k}_1,\vec{k}_2,\vec{k}_3) \equiv U$.  Most
calculations were performed using 48 patches.  Furthermore we
calculate the flow of several static susceptibilities, as described
below.  In addition we consider the flow of the couplings $h_c
(\vec{k})$ of quasiparticles at different positions on the Fermi
surface to uniform static charge fields.  This allows us to analyze
which classes of coupling functions and which susceptibilities become
important at low $T$.  For a large parameter range, we observe a flow
to strong coupling, i.e. at sufficiently low temperature some
components of the coupling function $V_{T}
(\vec{k}_1,\vec{k}_2,\vec{k}_3)$ become larger than the bandwidth.
The approximations mentioned above fail when the couplings are too
large.  Therefore we stop the flow when the largest coupling exceeds a
high value larger than the bandwidth, e.g. $V _{T, \mathrm{max}}=18t$.
This defines a {\em characteristic temperature} $T^*$ of the flow to
strong coupling.

\section{Antiferromagnetic, $d$-wave pairing,  $d$-density waves and
  $d$-wave Pomeranchuk fluctuations}
In this section we introduce fermionic coupling terms
to order parameters or static external fields corresponding to
different symmetry breaking channels.  Within the RG treatment
described in the previous section, we then calculate the one-loop
renormalizations of these coupling terms and thus obtain information
on the growth of the corresponding fluctuations in the low-temperature
state. A divergence of one of these external couplings signals a
crossover to a region with a strong tendency to ordering which could
become an actual finite temperature phase transition when coupling in
a third spatial direction is added.

First we introduce the $d$-wave pairing field
$\Phi_{s,d\mathrm{-sc}}=-\Phi_{-s,d\mathrm{-sc}}$, which couples to
the electrons via  
\begin{equation}  
\Phi_{s,d\mathrm{-sc}}  \,  \sum_{\vec{k}} 
h_{d\mathrm{-sc}}(\vec{k})
  \, c^\dagger_{\vec{k},s}  c^\dagger_{-\vec{k},-s} \, . 
\label{phidwp} 
\end{equation}
with a coupling constant $h_{d\mathrm{-sc}}(\vec{k})$. As initial
condition at $T=T_0$ we assume a $d_{x^2-y^2}$ form factor $
h_{d\mathrm{-sc}}(\vec{k}) =\left( \cos k_x - \cos k_y \right) /
\sqrt{2}$. Like the other coupling constants defined below,
$h_{d\mathrm{-sc}}$ will develop a $\vec{k}$-dependence through the
perturbative corrections in the course of the temperature flow towards
lower $T$ and higher harmonics of the same representation of the point
group will be admixed.  Next we also define a ($s$-wave) spin-density
wave field $\vec{\Phi}_{\mathrm{AF}}$. Due to spin-rotational
invariance it is sufficient to consider staggered fields in the spin
quantization direction,
\begin{equation}  \Phi_{z,\mathrm{AF}} \, \sum_{\vec{k}} h_{
    \mathrm{AF}}(\vec{k}) \, \left(  
c^\dagger_{\vec{k}+\vec{Q},s}  c_{\vec{k},s} - c^\dagger_{\vec{k}+\vec{Q},-s}
c_{\vec{k},-s} \right) \, .  
\label{phisdw}
\end{equation}
Here the momentum transfer $\vec{Q} = (\pi,\pi)$ between created and
annihilated particles corresponds to an alternating field acting on
the electron spins on n.n. sites. The initial condition at high
temperatures is $ h_{ \mathrm{AF}}(\vec{k}) =1$.  The proposed
Pomeranchuk deformations also have $d_{x^2-y^2}$ symmetry, the
coupling term is given by
 \begin{equation} 
\lim_{\vec{q} \to 0}\,   \Phi_{d\mathrm{-P}}
   \sum_{\vec{k},s} h_{d\mathrm{-P}}(\vec{k}) \,
   c^\dagger_{\vec{k}+\vec{q},s}  c^\dagger_{\vec{k},s}  \, , 
\label{phidP}
\end{equation}
with initial condition $h_{d\mathrm{-P}}(\vec{k}) = \left( \cos k_x -
  \cos k_y \right) / \sqrt{2}$ at $T_0$. A non-zero $d$-wave
Pomeranchuk field $\Phi_{\mathrm{d-P}}$ lifts the degeneracy of the
saddle points at $(\pi,0)$ and $(0,\pi)$ and breaks the fourfold
symmetry of the kinetic energy.  The $d$-density wave fluctuations
couple to particle-hole pairs with momentum transfer $\vec{Q}$,
\begin{equation}
\Phi_{d\mathrm{-DW}} \, \sum_{\vec{k},s}   h_{d\mathrm{-dw}}(\vec{k})\,  
c^\dagger_{\vec{k}+\vec{Q},s}  c_{\vec{k},s}\,  
\label{phiddw}
\end{equation}
with initial conditions chosen as \[  h_{d\mathrm{-DW}}(\vec{k}) =
\left( \cos  k_x - \cos 
k_y \right) / \sqrt{2} \] at the initial temperature $T_0$. 
In the one-loop $T$-flow  RG treatment~\cite{honerkampfm,salmhofer} 
the coupling constants $h$ are then renormalized according to 
\begin{eqnarray}
 \frac{d}{dT} {h}_{d\mathrm{-sc}}(\vec{k}) &=& -
\frac{1}{N_L} \,\sum_{\vec{k}'}  h_{d\mathrm{-sc}} (\vec{k}') \, V_T
(\vec{k},-\vec{k}, \vec{k}') \nonumber  \\ 
&& \cdot \, \frac{d}{dT}
L_{\mathrm{PP}} (\vec{k}',-\vec{k}') \, , \\
\frac{d}{dT} {h}_{\mathrm{AF}} (\vec{k}) &=& -
\frac{1}{N_L} \sum_{\vec{k}'} h_{\mathrm{AF}}  (\vec{k}')\, V_T
(\vec{k},\vec{k}', \vec{k}'+\vec{Q} )  
\nonumber  \\ 
&& \cdot \, \frac{d}{dT} L_{\mathrm{PH}}
(\vec{k}',\vec{k}'+\vec{Q}) \, , \\
\frac{d}{dT} {h}_{d\mathrm{-DW}}(\vec{k}) &=& -
\frac{1}{N_L} \sum_{\vec{k}'}h_{d\mathrm{-DW}}  (\vec{k}')
\frac{d}{dT} L_{\mathrm{PH}} 
(\vec{k}',\vec{k}'-\vec{Q}) \, \nonumber \\ 
&& \hspace{-1cm} \cdot \left[
V_T (\vec{k},\vec{k}', \vec{k}'+\vec{Q} ) - 2V_T (\vec{k},\vec{k}',
\vec{k}+\vec{Q} ) \right]  , \\
\frac{d}{dT} {h}_{d\mathrm{-P}}(\vec{k}) &=& -
\frac{1}{N_L} \sum_{\vec{k}'}  h_{d\mathrm{-P}}  (\vec{k}') \,
\frac{d}{dT} L_{\mathrm{PH}} 
(\vec{k}',\vec{k}') \nonumber  \\ 
&& \cdot  \left[
V_T (\vec{k},\vec{k}', \vec{k}' ) - 2V_T (\vec{k},\vec{k}',
\vec{k})  \right] 
 \,  .
\end{eqnarray}
The expressions on the right hand side are evaluated in absence of the
external fields. $N_L$ denotes the number of lattice sites and the
particle-particle and particle-hole loops are given by
\[ L_{\mathrm{PP}} (\vec{k},\vec{k}') = T \sum_{i\omega_n} 
\frac{1}{i\omega_n  - \epsilon_{\vec{k}}} \frac{1}{-i\omega_n  -
\epsilon_{\vec{k}'}} \] and
\[ L_{\mathrm{PH}} (\vec{k},\vec{k}') = T \sum_{i\omega_n} 
\frac{1}{i\omega_n  - \epsilon_{\vec{k}}} \frac{1}{i\omega_n  -
\epsilon_{\vec{k}'}} \, , \]
where $i\omega_n$ denotes the fermionic Matsubara frequencies
$\omega_n = \pi T (2n+1)$ with $n=1,2,\dots$.
The susceptibilities for the various channels are also renormalized
by temperature derivatives of bubble diagrams with the corresponding
 couplings at the vertices. 
For example for the $d$-wave pairing susceptibility we obtain 
\begin{eqnarray}  
\frac{d}{dT} {\chi}_{d\mathrm{-sc}} &=& \frac{1}{N_L} \sum_{\vec{k}}  
h_{d\mathrm{-sc}}(\vec{k})  \nonumber \\
 && \qquad \quad \cdot \,
 \frac{d}{dT} L_{\mathrm{PP}} (\vec{k},-\vec{k})\,  
h_{d\mathrm{-sc}}(-\vec{k})  \label{susflow}
\, . \end{eqnarray}  
Analogous expressions hold for the other susceptibilities. As initial 
condition we assume that all susceptibilities are zero at the initial
high temperature, $T_0$.

\section{Results for the 2D Hubbard model}
Here we describe our numerical results obtained with the $N$-patch
implementation of the one-loop $T$-flow RG scheme with $N=48$ patches.
For all cases discussed below, the RG flow goes to strong coupling.
This means that some components of the coupling function become larger
than the perturbative range when the temperature is lowered far
enough.  We define a characteristic temperature $T^*$ for the flow to
strong coupling determined as the temperature when the largest
coupling reaches $V_{\mathrm{max}} =18t$.

\subsection{The case $t'=0$}
\begin{figure}
\begin{center} 
\includegraphics[width=.49\textwidth]{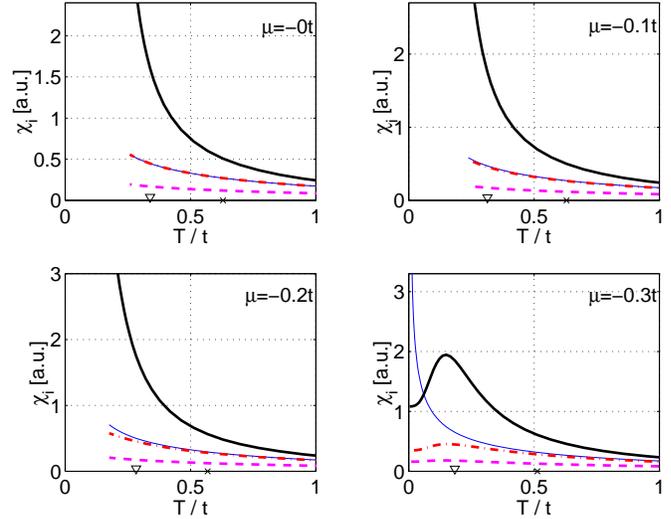}
\end{center} 
\caption{Temperature flow of AF (thick solid line), $d$-wave $sc$
  pairing (thin solid line), $d$-density wave (dashed-dotted line) and
  $d$-Pomeranchuk susceptibilities for different values of the
  chemical potential and $t'=0$. The cross and the triangle at the $T$
  axis denote the temperatures where the largest couplings exceed $5t$
  and $10t$. Half filling corresponds to $\mu=0$.}
\label{sus48tp0}
\end{figure}

First let us discuss results for vanishing next-nearest neighbor hopping,
$t'=0$. The flow of the susceptibilities is summarized in
Fig. \ref{sus48tp0}.

For the square Fermi surface at half band filling, i.e. $\mu=0$, we
find that the AF susceptibility is the leading divergent
susceptibility, consistent with the expected AF ordered ground state.
We also find weaker growing $d$-wave pairing and $d$-density wave
susceptibilities. At $\mu=0$, these two channels have exactly the same
temperature dependence due to the particle-hole symmetry; $\epsilon
(\vec{k}) = - \epsilon (\vec{k} + \vec{Q})$.  For finite doping, i.e.
$\mu \not= 0$, the flow splits this degeneracy and the $d$-wave
pairing generally grows more strongly than the $d$-density wave
channel.  As the electron density is further reduced below
half-filling, the nesting in the $(\pi,\pi)$ particle-hole channel is
increasingly weakened. Correspondingly, at a critical doping that
depends on the interaction strength, the growth in the AF and
$d$-density wave channels get cut off at low $T$ and the $d$-wave
pairing is the only divergent channel that remains. This crossover
from the AF dominated flow to the $d$-wave pairing dominated
instability for $t'=0$ is very similar to the results of Zanchi
and Schulz~\cite{zanchi} and Halboth and Metzner~\cite{halboth} who
used a $N$-patch scheme with flowing IR cutoff. Within the latter
schemes it is difficult to compare the flow of small-$\vec{q}$
particle-hole susceptibilities, e.g.  Pomeranchuk FS deformations,
with the flow of AF or superconducting susceptibilities. 
There is no such difficulty in the temperature flow scheme.
Thus we can directly
compare the Pomeranchuk channel with the above-mentioned other
tendencies. The main conclusion arising from the data shown in
Fig.~\ref{sus48tp0} is that the $d$-wave Pomeranchuk tendencies grow
somewhat to lower temperatures, but apparently they do not represent
the leading instability for the $t'=0$ Hubbard model with only a local
onsite repulsion as in initial interaction.
In this paper we do not give a systematic analysis how the leading
instability may change when the initial interaction is varied, e.g. by
including a n.n.n.  repulsion $V$ or a spin exchange coupling $J$.
This question was addressed by Binz et al.~\cite{binz} for the
half-filled case and $U \to 0$. These authors found a $d$-density wave
regime for sufficiently large positive n.n. Heisenberg coupling $J$
between an AF regime at small n.n.  repulsion $V$ and $s$-charge
density wave regime at $4V>U$.  We find a similar crossover between
the two latter regimes for $U \ge 0.3t$.  However a RG flow where the
$d$-density wave susceptibility grows most strongly was only found for a
limited parameter region around $J \ge 0.7 U$ and $4V\approx 0.7U$.
   
\subsection{Moderate n.n.n. hopping $t'=-0.25t$}
\begin{figure}
\begin{center} 
\includegraphics[width=.4\textwidth]{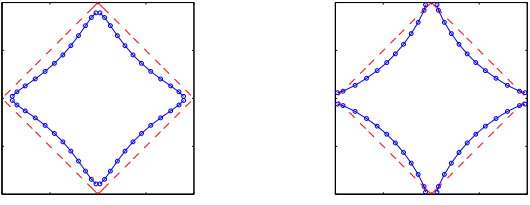}
\end{center} 
\caption{Non-interacting Fermi surfaces with $t'=-0.25t$ for $\mu=-1.05t$
  (left plot, $\langle n \rangle \approx 0.75$ per site) and
  $\mu=-0.95t$ (right plot, $\langle n \rangle \approx 0.82$ per
  site). The dots denote the location of the 48 points for which the
  flow of the coupling function is evaluated. The dashed square
  denotes the so-called Umklapp surface.}
\label{fs105_95}
\end{figure}

Next we focus on the case of n.n.n. neighbor hopping $t'=-0.25t$ and
band fillings $n<1$ for which the FS passes close to the saddle points.
Upon increasing the particle density through the van Hove filling at
which $\mu$ lies exactly at the van Hove singularity, i.e. $\mu=4t'$,
the FS starts to intersect the Umklapp surface which connects the
$(\pi , 0)$ and $(0,\pi )$ points with straight lines. As we will
show, this gives rise to a remarkable qualitative change in the flow.
Two typical shapes of the the non-interacting Fermi surfaces for
densities below and above the van Hove filling are shown in Fig.
\ref{fs105_95}.

\begin{figure}
\begin{center} 
\includegraphics[width=.49\textwidth]{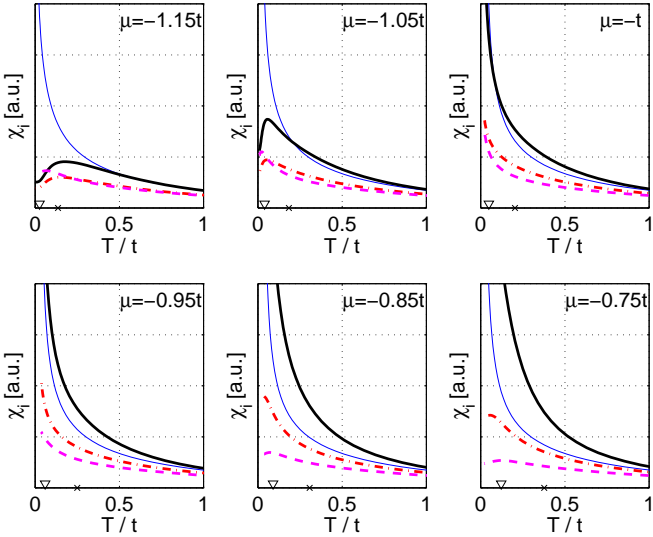}
\end{center} 
\caption{Temperature flow of AF (thin solid line), $d$-wave pairing
  (thick solid line), $d$-density wave (dashed-dotted line) and
  $d$-Pomeranchuk susceptibilities for different values of the
  chemical potential and $t'=-0.25t$. The cross and the triangle at
  the $T$ axis denote the temperatures where the largest couplings
  exceed $5t$ and $10t$. Band fillings less than the
  saddle point filling correspond to $\mu<-t$. $\langle n \rangle \approx 0.93$ per site for $\mu=-0.75t$. }
\label{sus96tp25}
\end{figure}

In Fig. \ref{sus96tp25} we show the evolution of $d$-wave pairing, AF,
$d$-density wave and $d$-wave Pomeranchuk susceptibilities for
different values of the chemical potential as the temperature is
lowered.  In the first upper two plots for which $\mu<4t'$, we observe
that only the $d$-wave pairing susceptibility grows rapidly at low
temperatures $T \to T^* >0$.  The other susceptibilities reach a
maximum at a $T_{\mathrm{max}} > T^*$ and decrease again as $T \to
T^*$.  For the next two plots the FS intersects the Umklapp surface
close to the saddle points and the flow is qualitatively different:
now all four susceptibilities grow as $T\to T^*$. As described in
Ref.~\cite{honerkamp}, the flows of AF and $d$-wave pairing
susceptibilities are very similar for a broad density region above the
saddle point filling and it appears to be implausible that the strong
coupling state is a simple symmetry-broken state corresponding to a
single type of ordering. More than that, the $d$-density wave and
$d$-wave Pomeranchuk susceptibilities also grow as $T\to T^*$. But for
our parameters these two channels do not constitute the leading
instabilities. Nevertheless they seem to be part of a common mechanism
arising from the special location of the FS close to saddle points
which causes mutual reinforcement of different channels, most
prominently AF and $d$-wave pairing described in
Ref.~\cite{honerkamp}. The main reason for this effect is that the
scattering processes connecting the two inequivalent saddle point
regions involve particle pairs with small total momentum that scatter
with wavevector transfers close to $(\pi, \pi)$. Therefore these
scattering processes are strongly driven by both the Cooper and the
$(\pi, \pi)$ particle-hole channel and this causes a mutual
reinforcement of the fluctuations generated by these channels.  Thus
it is interesting to speculate that the strong coupling state will
embody all these channels as short range correlations, instead of
selecting a single ordering channel and suppressing the others.
Indeed, as discussed in the concluding section, examples for such ground
states are known from ladder systems.  Furthermore, other numerical
investigations of the doped 2D $t$-$J$ model showed the existence of
$d$-density wave correlations in Gutzwiller-projected $d$-wave pairing
variational wave-functions ~\cite{ivanov} and in the lowest energy
$d$-wave paired states in exact diagonalization on a 32-site
cluster~\cite{leung}.

\begin{figure}
\begin{center} 
\includegraphics[width=.49\textwidth]{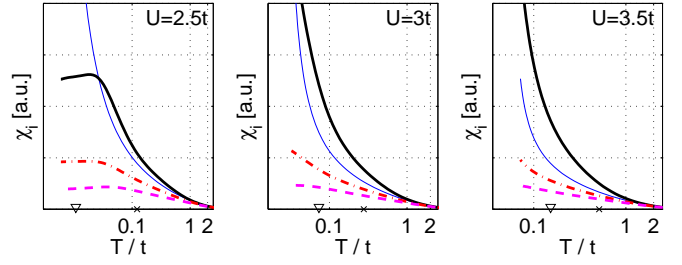}
\end{center} 
\caption{Temperature flow of AF (thin solid line), $d$-wave pairing
  (thick solid line), $d$-density wave (dashed-dotted line) and
  $d$-Pomeranchuk susceptibilities for different values of the initial
  interaction $U$ and $\mu=-0.92t$ at $t'=-0.25t$. The cross and the
  triangle at the $T$ axis denote again the temperatures where the
  largest couplings exceed $5t$ and $10t$.}
\label{sus92comp}
\end{figure}

We note that if we continue to follow the one-loop RG flow for the 2D
system which ignores selfenergy corrections, the common growth of the
different channels does not continue to arbitrarily low temperatures.
Typically a decoupling takes place once the overlap between the
channels becomes too small due to the restriction of the relevant
scattering processes at low temperatures to thinner and thinner
neighborhoods of the Fermi surface. Signs for this are visible for
example in the flow of $d$-density wave susceptibility in the last
plot of Fig.~\ref{sus96tp25}. However we are convinced that one should
concentrate on the flow down to temperatures where the growing interactions
reach values comparable to the bandwidth. This is also the scale where
the corrections to the one-loop RG equations such as one- and 
two-loop selfenergy effects start to become 
non-negligible~\cite{honerkamp,honerkamped}. In the saddle
point regime described above the decoupling of the different channels
only occurs at lower temperatures where the one-loop flow without
selfenergy corrections has lost its validity as the couplings are much
larger than the bandwidth. Obviously, the density range where this
common flow of several channels occurs becomes larger with increasing
interaction strengths $U$. 
The reason is that at higher temperature
scales a wider shell around the FS participates in the RG flow and the
overlaps between the scattering processes driving the individual
fluctuations are larger. This is illustrated in Fig.~\ref{sus92comp},
where we plot the flow of the susceptibilities for the same chemical
potential but for different initial interactions $U$. While for
$U=2.5t$ the flows of AF, $d$-density wave and $d$-wave Pomeranchuk
susceptibilities go through a maximum before the couplings exceed the
bandwidth, for $U\ge 3t$ their flow grows monotonically over the range
where the one-loop flow applies. Thus we expect that there is a nonzero 
minimal interaction strength above which the coupled saddle point flow 
determines the ground state properties. For smaller interactions 
the low energy physics may be dominated by a single channel.  

\begin{figure}
\begin{center} 
\includegraphics[width=.49\textwidth]{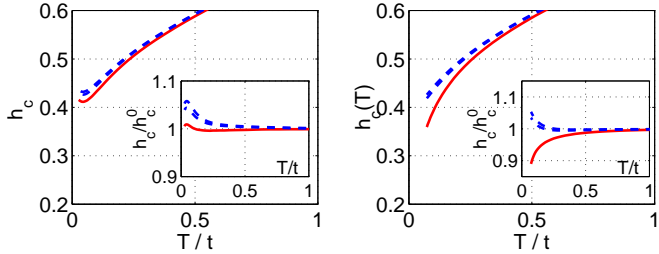}
\end{center} 
\caption{Temperature flow of couplings $h_c(\vec{k}_F)$ of quasiparticles on
  different points $\vec{k}_F$ on the FS to uniform static charge
  fields.  The insets show the flow normalized to the RPA values for
  the corresponding temperatures. The solid (dashed) lines correspond
  to $\vec{k}_F$ points close to the saddle points (BZ diagonals). The
  left plot is for $\mu=-1.1t$ (band filling less than the van Hove
  value, $d$-wave regime, band filling less than van Hove filling),
  and the right plot for $\mu=-0.95t$ (saddle point regime, band
  filling slightly larger than van Hove filling).}
\label{cctp25}
\end{figure}

With the new $T$-flow RG scheme we again find an anisotropic
suppression of the charge compressibility in the saddle point regime,
similar to our findings in Ref.~\cite{honerkamp}. This is illustrated
in Fig.~\ref{cctp25}, where we plot the $T$-flow of the couplings to
external static uniform charge fields, $h_c (\vec{k})$. These
quantities are calculated in a way analogous to the Pomeranchuk
coupling constants $h_{d\mathrm{-P}}(\vec{k})$, with the difference
that the $d$-wave form factor in Eq.~(\ref{phidP}) is replaced by
unity. In the left plot the density is less than the van Hove filling
corresponding to the pure $d$-wave regime and the charge couplings
$h_c (\vec{k})$ are suppressed uniformly around the FS. Moreover, as
shown in the inset, the RG flow of the interactions enhances the
charge couplings $h_c(\vec{k})$ with respect to isotropic RPA (random
phase approximation) values $h^0_c$ calculated without RG flow of the
interactions. In the right plot, we show the same quantities for
$\mu=-0.95t$, corresponding to the saddle point regime at band
fillings slightly larger than the van Hove value. Here the suppression
of the charge couplings is strongest in the FS region close to the
saddle points. Furthermore, compared to the RPA values, the charge
couplings for this FS region are strongly decreased at low $T$ by the
RG flow of the interactions, while they increase near to the BZ
diagonals. We emphasize that these tendencies are rather weak because
we do not let the interactions become too strong.  Nevertheless, our
findings with the new $T$-flow RG scheme are again consistent with the
hypothesis of a Fermi surface truncation~\cite{furukawa,honerkamp} at
the saddle points where the scattering processes grow most strongly,
with remaining FS arcs around the BZ diagonals. We believe that the
effects observed as tendencies in the weak-coupling approach will be
strongly amplified in a full strong coupling description. We add that
a similar suppression of the $\vec{k}$-space resolved compressibility
at the saddle points is observed within a Landau-Fermi liquid
framework~\cite{frigeri} and was also found in quantum Monte Carlo
calculations for the half filled Hubbard model by Otsuka et
al.~\cite{morita}.

Lastly we comment on the behavior of the $T$-flow RG scheme as the
density is increased towards half filling. 
The saddle points move to larger negative band energies and the coupling between the channels becomes less effective (see middle and right lower plots in Fig. \ref{sus96tp25}). 
The AF susceptibility grows most strongly at higher $T$.
At lower temperatures the $T$-flow RG scheme
exhibits a stronger sensitivity to the imperfect nesting introduced by
the n.n.n. hopping term $t'$ than in the IR-flow RG scheme we used
previously~\cite{honerkamp}. This shows up in a saturation of the AF
susceptibility flow for weaker initial interactions at low 
$T$ before the couplings have reached values larger than the bandwidth. At
half-filling an initial value of $U\geq 3.5t$ is required 
to obtain the dominant AF phase found in Ref.~\cite{honerkamp}. 

\section{Conclusions and comparison with the two-leg Hubbard ladder}
We have investigated the flow of $d$-wave pairing, spin density wave,
$d$-density wave and $d$-wave Pomeranchuk susceptibilities in the
two-dimensional Hubbard model near the van Hove filling. For a n.n.n.
hopping $t'=-0.25t$ we found two distinct regimes.  For densities well
away from the van Hove density, only a single susceptibility diverges
at low $T$. For a range of band fillings $n<1$ around, or somewhat
above, the van Hove filling, whose width depends on the strength of
the initial interaction, several susceptibilities grow together when
the interactions flow to strong coupling. In particular, from the
interplay between $d$-wave pairing and AF tendencies near the saddle
points, also described in Ref.~\cite{honerkamp}, we conclude that the
strong coupling phase, in the saddle point regions, is not given by a
dominance of a single ordering such as superconducting, magnetic or
charge ordering.  Furthermore we again find tendencies towards
incompressibility of the FS Pomeranchuk regions near the saddle points
which can be interpreted as indication for a FS truncation. Although
the $d$-density wave and $d$-wave FS deformation modes get enhanced at
lower temperatures, but do not represent the leading instabilities for
the $t$-$t'$ Hubbard model with weak to moderate repulsive onsite
interaction. This renders a spontaneous symmetry breaking in these
channels unlikely. Nevertheless there may be significant short ranged
correlations of the respective types.

A similar RG flow to strong coupling is also found in the half-filled
two-leg Hubbard ladder. This quasi-one-dimensional system has been
extensively analyzed both from the weak-coupling
perspective~\cite{balents,lin,fisher} and by numerical methods for
stronger onsite repulsion~\cite{noack}. Through these studies it
became clear that the ground state for all $U$ is given by an {\em
  insulating spin liquid}, i.e. spin and charge fluctuations are
gapful and decay exponentially. The FS is fully truncated although no
symmetry breaking occurs.  

The weak-coupling analysis~\cite{balents,lin} typically starts with a
one-loop RG scheme which selects the relevant couplings. Subsequently
the effective Hamiltonian, composed of the kinetic energy and the
relevant couplings, is bosonized and the properties of spin and charge
excitations can be deduced. Our comparison with the 2D Hubbard model
focuses on only this one-loop RG step.  In the one-loop g-ology for
the half-filled two-leg ladder, one finds seven relevant coupling
constants $g_i$, $i=1, \dots, 7$ which describe scattering processes
between the 4 Fermi points, 2 in the even parity band and 2 in the odd
band.  In the RG flow with decreasing infrared cut-off $\Lambda$ and
initial condition $g_i (\Lambda_0) =U$ these couplings constants
diverge asymptotically together as
\begin{equation} 
g_i(\Lambda) = \frac{g^0_i}{\log(\Lambda/\Lambda_c )} \, , 
\label{rgansatz} 
\end{equation}
when $\Lambda$ is reduced towards a finite critical scale $\Lambda_c$.
The $g^0_i$ reach fixed ratios in the weak interaction limit.  Due to
the divergence of the coupling functions, the one-loop
renormalizations drive several couplings to external fields to
infinity. Like in the 2D system several types of fluctuations are
driven by common scattering processes and mutually reinforce each
other.  As mentioned in the last section, 
in the 2D model we need a finite initial interaction strength $U$ to observe
the coupled flow of several channels for a reasonably wide parameter region. 
In contrast with that in the two-leg ladder the mutual reinforcement is present at all energy scales and occurs for arbitrarily weak interactions. 
More precisely, with the asymptotic flow given by
Eq.~(\ref{rgansatz}) we find that in the half-filled ladder 
AF, $d$-wave pairing, $d$-density
wave and spin density wave couplings grow with the same
strength\footnote{This is also true for the 2D model restricted to the
  saddle point regions, see Ref.~\cite{binz}}.  Equivalently, if we
write down the effective Hamiltonian of the system close to the
critical scale, the effective coupling constants for mean-field
decouplings in $d$-wave pairing, AF and $d$-density wave channel reach
the same absolute value in the asymptotic flow very close to the
instability.  Since the effective densities of states for the three
types of mean-field decouplings are the same, their potential ground
state energies are degenerate as well.  Nonetheless, as we know from
the bosonization~\cite{lin} and numerical studies~\cite{noack}, the
ground state does not exhibit (quasi-)long-range order of any type and
the superconducting and magnetic correlations which seem to diverge in
the one-loop treatment remain strictly short-ranged. 

As another similarity between the 2D system and the two-leg ladder we
mention that also in the ladder the forward scattering amplitudes
calculated within the RG scheme exhibit a $d$-wave-type structure with
different signs for the forward scattering onto the same band and onto
the other band.  However a shift of the Fermi points induced by the
flow of the forward scattering processes was found in a weak-coupling
RG treatment only for the case of unequal Fermi velocities in the two
bands~\cite{louis}.

Finally we add that in the 2D Hubbard model with $t'<0$ for band
fillings $n>1$ corresponding to electron-doped case the situation is
different  from the hole-doped case described above\cite{honerkamped}.
For $n>1$ the $(\pi,0)$ and $(0,\pi )$ regions of the BZ, that are
crucial for the channel coupling leading to the combined flow in the
saddle point regime similar to the two-leg ladder flow, are at larger
negative band energies and exert only little influence on the low
temperature physics. Correspondingly, upon moving away from half
filling by increasing $n>1$ and suitable parameters, the flow to
strong coupling rapidly changes from an AF regime with high $T^*$ to a
$d$-wave pairing regime with low $T^*$ without the occurrence of an
intermediate saddle point regime.

In summary, the results of the recently developed $T$-flow one loop RG
scheme applied to the $t$-$t'$-$U$ Hubbard model in 2D, are
qualitatively consistent with those we found earlier using an IR-flow
RG scheme. The $T$-flow RG scheme has the advantage that the
competition with long wavelength instabilities such as a $d$-wave FS
Pomeranchuk distortion can be analyzed in an unbiased way. Our results
show that this susceptibility and also the $d$-density wave
susceptibility are never the dominant ones in a Hubbard with just an
onsite repulsion, $U$ although for densities in the saddle point
regime they appear as divergent susceptibilities together with those
towards $d$-wave pairing and AF order. These results agree with the
proposal we made earlier~\cite{honerkamp} regarding the nature of the
strong coupling phase that appears at temperature $T\leq
T^*$. Based on the close analogy with the one loop RG results for the
two leg ladder at half-filling, we proposed that at $T\approx T^*$ a
crossover occurs to a strong coupling phase with a FS which is
partially truncated in the vicinity of the saddle points by the
formation of an insulating spin liquid which in turn is not
characterized by any, even quasi, long range order. The remaining FS
arcs centered on the BZ diagonals are coupled in the Cooper channel
to the insulating spin liquid and this coupling is the origin of a
transition to $d$-wave superconductivity at a lower $T<T^*$. Note we
find that $T^*$ is a strong function of the density and increases from
zero with increasing density as the van Hove filling is crossed. This
behavior of $T^*$ is, we argue, consistent with the critical density and
rising temperature scale found by Tallon and Loram~\cite{tallon}.

Of course our one-loop approach does not yield a controlled theory of
the strongly coupled phase below the characteristic temperature $T^*$
and the second step in the weak-coupling analysis of the exemplary
ladder systems, the bosonization of the effective Hamiltonian, cannot
be performed in the 2D case. Thus we cannot prove the properties of
the 2D strong coupling phase for the time being. More powerful methods
need to be developed in order to %
establish the nature of the strong coupling phase at temperatures
$T<T^*$, to test our proposal that it is closely analogous to that of
the 2-leg ladder and also to examine possible connections to the gauge
theories which proceed directly from strong coupling and also predict
strong fluctuations coexisting in several channels such as the
$d$-wave pairing and $d$-density wave channels~\cite{palee}.
\\[2mm]

We thank M. Sigrist, M. Troyer, D. Poilblanc, W. Metzner, and C. Nayak
for helpful discussions. C.H. acknowledges support by the Swiss
National Science Foundation (SNF) and the Emmy-Noether program of the
Deutsche Forschungsgemeinschaft (DFG). The numerical calculations
were performed on the {\em Asgard} Beowulf cluster of ETH Z\"urich.

\end{document}